\documentclass[aip,
amsmath,amssymb,
reprint,]{revtex4-1}

\usepackage{graphicx}\usepackage{dcolumn}\usepackage{bm}

\usepackage[utf8]{inputenc}
\usepackage[T1]{fontenc}
\usepackage{mathptmx}
\usepackage{etoolbox}
\usepackage{float}
\usepackage{placeins}

\makeatletter
\def\@email#1#2{\endgroup
 \patchcmd{\titleblock@produce}
  {\frontmatter@RRAPformat}
  {\frontmatter@RRAPformat{\produce@RRAP{*#1\href{mailto:#2}{#2}}}\frontmatter@RRAPformat}
  {}{}
}\makeatother
\begin{document}

\preprint{AIP/123-QED}

\title[Controlling pair dynamics of rotating magnetic microparticles through radial and transverse interactions]{Controlling pair dynamics of rotating magnetic microparticles through radial and transverse interactions}
\author{Dongfang Fu}
\affiliation{State Key Laboratory for Turbulence and Complex Systems, School of Mechanics and Engineering Science,
Peking University, Beijing 100871, China }

\author{Leilei Wang}
\affiliation{State Key Laboratory of Nonlinear Mechanics, Beijing Key Laboratory of Engineered Construction and Mechanobiology,
Institute of Mechanics, Chinese Academy of Sciences, Beijing 100190, China}

\author{Kailai Wang}
\affiliation{State Key Laboratory of Nonlinear Mechanics, Beijing Key Laboratory of Engineered Construction and Mechanobiology,
Institute of Mechanics, Chinese Academy of Sciences, Beijing 100190, China}\affiliation{School of Building Services Science and Engineering, Xi’an University of Architecture and Technology, Xi’an 710055, China}

\author{Xu Zheng}
\affiliation{State Key Laboratory of Nonlinear Mechanics, Beijing Key Laboratory of Engineered Construction and Mechanobiology,
Institute of Mechanics, Chinese Academy of Sciences, Beijing 100190, China}

\author{Zaiyi Shen}\email{zaiyi.shen@pku.edu.cn}
\affiliation{State Key Laboratory for Turbulence and Complex Systems, School of Mechanics and Engineering Science,
Peking University, Beijing 100871, China }

\date{\today}

\begin{abstract}
Rotating magnetic microparticles are building blocks for field-driven assembly and microrobotic control. As the elementary interaction rule for larger assemblies, pair motion in these systems is governed not only by magnetic forcing, but also by hydrodynamic coupling and other long-range interactions. Here we develop a reduced framework for two synchronized rotating magnetic particles by resolving the interactions into radial components that change the interparticle distance and transverse components that rotate the line of centers. The competition between magnetic dipolar interaction, additional radial repulsion, and rotation-induced transverse coupling selects three pair-motion modes: rigid-body rotation, contact-separation rotation, and irreversible separation. We derive transition criteria for the rigid-body state, reversed orbital motion, and the separation boundary, and obtain an asymptotic solution for the separation dynamics. Lattice Boltzmann simulations of particles rotating near a wall provide a hydrodynamic realization of the model, in which inertial secondary flow generates radial repulsion and rotational flow produces transverse coupling. The resulting phase diagram in the physical $(\mathrm{Re},C_m)$ plane is consistent with the reduced-model predictions. These results provide design rules for programming elementary pair interactions in rotating magnetic-particle systems and may help guide the control of microrobotic assemblies.
\end{abstract}

\maketitle

Magnetically driven micro and nanoparticles provide a versatile platform for untethered actuation, field-driven assembly, and microrobotic control.\cite{Nelson2010,Li2017,Ceylan2017,Peyer2013} Recent progress has shifted the focus from single-particle actuation toward coordinated multi-particle motion, reconfigurable assemblies, and swarm-level control, where the collective response is determined not only by the imposed magnetic field but also by particle-particle interactions and hydrodynamic coupling.\cite{Abbott2020,Yang2021,Xie2019,Yu2019} In such systems, pair dynamics provides the elementary interaction rule for larger assemblies, controlling local binding, rearrangement, and motion transmission.~\cite{Grzybowski2000,Grzybowski2002,Tierno2008}

Magnetic dipolar interactions can produce bound pairs, chains, rotating clusters, and reconfigurable structures under static or time-dependent fields.\cite{Grzybowski2000,Grzybowski2002,Tierno2008,Klapp2016,Snezhko2016,Coughlan2016,Coughlan2017,SpataforaSalazar2024} At the colloidal scale, however, pair motion is rarely governed by magnetic interactions alone. Depending on the driving protocol and material platform, magnetic forcing can compete with hydrodynamic coupling, electrostatic interactions, capillary forces, phoretic effects, or other field-mediated interactions.\cite{Ristenpart2004,Ma2015,Patteson2016,LiuSharifiMoodStebe2018,Palacci2013,AlHarraq2022,Kryuchkov2019} For a two-particle system, these effects can be naturally decomposed in the relative coordinate frame into radial and transverse components that govern the interparticle distance and the rotation of the line of centers, respectively.

Although previous pair-level studies have revealed rich interaction mechanisms, predicting pair motion remains challenging when magnetic attraction competes simultaneously with radial repulsion and transverse coupling. Existing work has addressed dipole-coupled spin dynamics, colloidal-pair relaxation, hydrodynamic bound states, and collective rotation in magnetic assemblies.~\cite{Laroze2008,Coughlan2016,Coughlan2017,MartinezPedrero2015,Tierno2010,MartinezPedrero2018,Belovs2019} Yet a compact pair-level framework that relates competing radial and transverse interactions to motion-mode selection and transition criteria is still lacking.

In this Letter, we develop such a reduced framework for two synchronized rotating magnetic microparticles. The model resolves the pair dynamics into radial motion of the interparticle distance and angular motion of the line of centers, with time-dependent magnetic dipolar interaction, additional radial repulsion, and rotation-induced transverse coupling as the leading ingredients. We show that their competition selects three characteristic modes: rigid-body rotation, contact-separation rotation, and irreversible separation. We derive the transition criteria for these modes, identify the threshold for reversed orbital motion within the contact-separation state, and obtain an asymptotic solution for the separation dynamics. Lattice Boltzmann simulations of two particles rotating near a wall then provide a hydrodynamic realization of the reduced model, where finite-inertia secondary flow generates radial repulsion and rotational flow produces transverse coupling. Together, the model and simulations establish pair-level design rules for controlling rotating magnetic microparticle systems.

\begin{figure}
\includegraphics[width=1\linewidth]{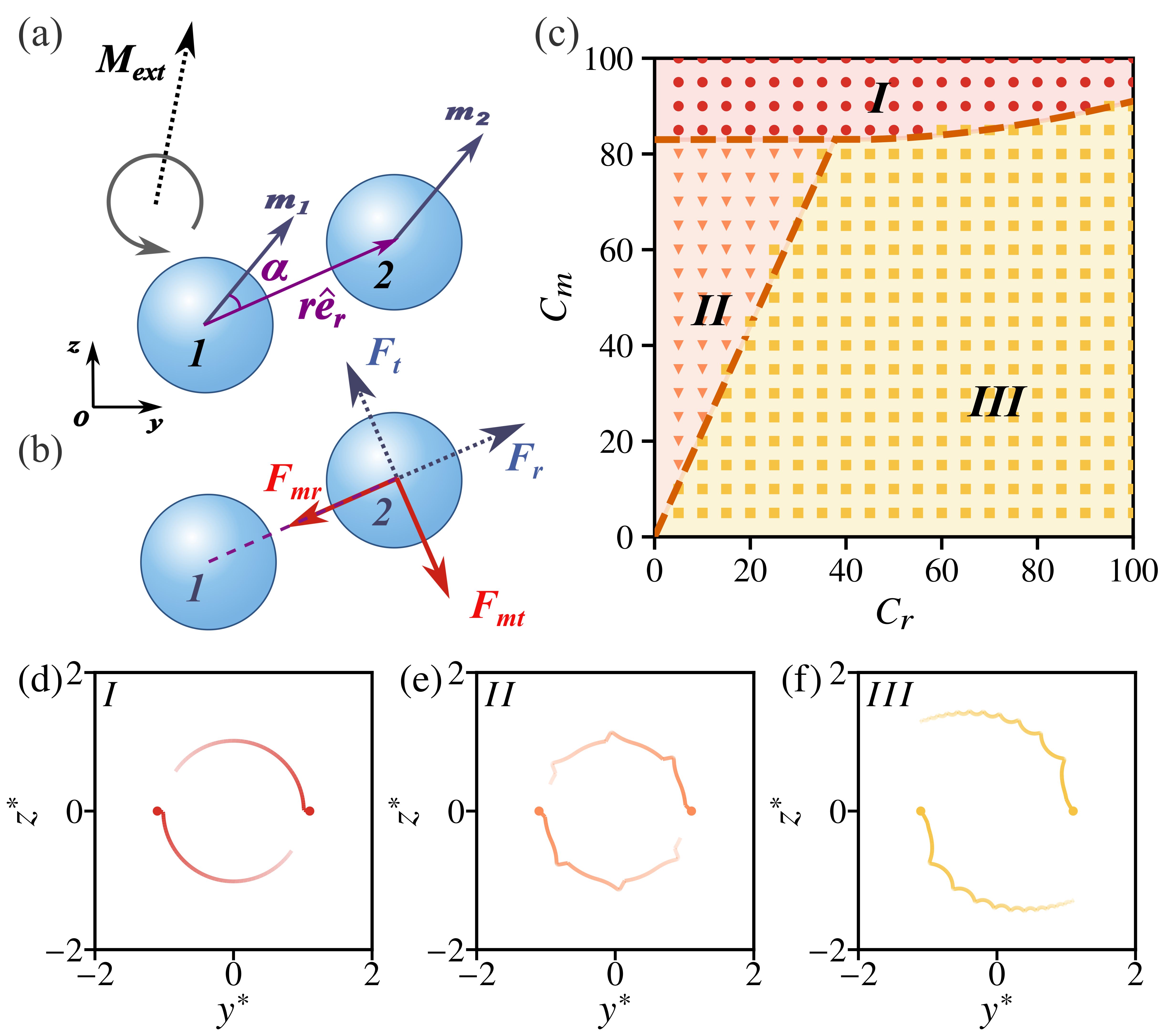}
\caption{\label{fig1}
Pair dynamics of two rotating magnetic microparticles in the reduced model.
(a) Schematic of the magnetic dipole model.
(b) Decomposition of the interactions into magnetic components and additional radial and transverse interactions.
(c) Phase diagram of pair dynamics in the $(C_r,C_m)$ parameter space for $p=3$ and $q=5$. \emph{Mode~I}: rigid-body rotation; \emph{Mode~II}: contact-separation rotation; \emph{Mode~III}: irreversible separation.
(d)–(f) Representative particle trajectories in the three modes. Solid circles indicate the initial particle positions.
}
\end{figure}

We consider two identical spherical magnetic microparticles of radius $R$, driven by a rotating external magnetic field. The magnetic interaction between the particles is modeled using the point-dipole approximation, with a permanent magnetic moment assigned to the center of each particle. The particle centers are restricted to the $yz$ plane, and their magnetic moments rotate in this plane at angular frequency $\omega$, about the wall-normal $x$ axis [Fig.~\ref{fig1}(a)]. The applied field is sufficiently strong to keep the magnetic moments synchronized with the external rotation. In this regime, the rotation of the particle is prescribed by the field, and only the translational dynamics needs to be solved. The magnetic force exerted by dipole $i$ on dipole $j$ is
\begin{equation}
\begin{aligned}
\boldsymbol{F}_{m,ij}
&= \frac{3\mu_{0}}{4\pi r^{4}}
\Big[
(\boldsymbol{m}_{j} \cdot \hat{\boldsymbol{e}}_{r}) \boldsymbol{m}_{i}
+(\boldsymbol{m}_{i} \cdot \hat{\boldsymbol{e}}_{r}) \boldsymbol{m}_{j} \\
&\qquad
-5(\boldsymbol{m}_{i}\cdot\hat{\boldsymbol{e}}_{r})
  (\boldsymbol{m}_{j}\cdot\hat{\boldsymbol{e}}_{r})
  \hat{\boldsymbol{e}}_{r}
+(\boldsymbol{m}_{i}\cdot\boldsymbol{m}_{j}) \hat{\boldsymbol{e}}_{r}
\Big] ,
\label{eq:mag}
\end{aligned}
\end{equation}
where $\mu_0$ is the vacuum permeability, $\boldsymbol{m}_i$ and $\boldsymbol{m}_j$ are the magnetic moments of the two dipoles, $r$ is the center-to-center distance between the particles, and $\hat{\boldsymbol{e}}_{r}$ is the unit vector directed from dipole $i$ to dipole $j$.

We resolve all interactions into radial and transverse components, along and perpendicular to the line of centers [Fig.~\ref{fig1}(b)]. The radial component changes the interparticle distance, whereas the transverse component rotates the line of centers and deflects the pair trajectory. For two synchronized dipoles with equal magnetic moment magnitude $m$, the magnetic force gives a radial term $\boldsymbol{F_{mr}} \sim -[1+3\cos(2\alpha)]/r^4$ and a transverse term $\boldsymbol{F}_{mt} \sim \sin(2\alpha)/r^4$, where $\alpha$ is the angle between the magnetic moment and the line of centers [Fig.~\ref{fig1}(a)].

In addition to the magnetic dipolar interaction, we include two effective interactions commonly present in driven magnetic-particle systems [Fig.~\ref{fig1}(b)]. The radial interaction $\boldsymbol{F}_r$ is introduced phenomenologically to represent an additional long-range repulsion beyond the magnetic dipolar interaction. Its physical origin may vary with the system, including electrostatic, colloidal, capillary, or inertial hydrodynamic effects~\cite{Kryuchkov2019,Patteson2016,LiuSharifiMoodStebe2018,ShenLintuvuori2020}. The transverse interaction $\boldsymbol{F}_t$ is associated with the azimuthal flow generated by particle rotation, which contributes to the transverse advection of the other particle~\cite{Bililign2022,ShenLintuvuori2023}.

After nondimensionalizing the distance and time as $r^\ast=r/R$ and $t^\ast=\omega t/2\pi$, respectively, the pair dynamics is written as
\begin{align}
\frac{dr^\ast}{dt^\ast}
= \frac{C_r}{r^{\ast p}}
-\frac{C_m\bigl(1+3\cos 2\alpha\bigr)}{r^{\ast 4}},
\label{eq:dry_r}
\\
\frac{d\alpha}{dt^\ast}
= 2\pi
-\frac{C_t}{r^{\ast q}}
-\frac{2C_m\sin 2\alpha}{r^{\ast 5}} .
\label{eq:dry_alpha}
\end{align}
Here $C_r$, $C_m$, and $C_t$ are nondimensional coefficients that set the strengths of the additional radial interaction $\boldsymbol{F}_r$, the magnetic dipolar interaction, and the rotation-induced transverse coupling $\boldsymbol{F}_t$, respectively. The exponent $p$ characterizes the spatial decay of the additional radial contribution, whereas $q$ characterizes the decay of the angular drift produced by the transverse coupling. Both exponents depend on the underlying physical mechanism. For example, hydrodynamic interactions generated by a rotating sphere near a solid surface at finite Reynolds number may give $p \approx 3$ and $q \approx 5$ (Fig.~S1 of the Supplementary Material). In the numerical integration, a short-range steric repulsion is added only to prevent particle overlap when $r^\ast<r_{s}^\ast=2.03$.

We study the pair dynamics in the $(C_r,C_m)$ parameter space while keeping $C_t\simeq 50$, a value corresponding to the transverse hydrodynamic coupling of particles rotating near a surface, as estimated from the fitting in Fig.~S1(b) of the Supplementary Material. The initial condition is set to $r^\ast(0)=2.2$ and $\alpha(0)=0$ in all simulations. Three typical modes of pair motion are observed as $C_r$ and $C_m$ are varied [Fig.~\ref{fig1}(c)].

When the magnetic interaction is sufficiently strong, the pair exhibits \emph{Mode~I}, a rigid-body rotation state: the particles remain in contact, $\alpha$ reaches a steady value, and the pair rotates synchronously with the field [Fig.~\ref{fig1}(d)]. When the magnetic interaction becomes weaker, phase locking is lost and $\alpha$ varies periodically. The radial magnetic interaction then alternates between attraction and repulsion, producing \emph{Mode~II}, in which the particles repeatedly contact and separate [Fig.~\ref{fig1}(e)]. For stronger additional radial repulsion, the particles separate without recontacting, giving \emph{Mode~III} [Fig.~\ref{fig1}(f)].

We further tested different values of the decay exponents $p$ and $q$ and found that they do not change the qualitative structure of the mode diagram (Fig.~S2 of the Supplementary Material). This robustness suggests that the three pair-dynamics modes are not tied to a specific microscopic origin of the additional interactions.

In the rigid-body rotation state, the magnetic attraction is sufficiently strong to overcome the additional radial repulsion, bringing the particles into contact ($\dot{r^\ast}<0$). Once contact is reached, the short-range steric interaction fixes the interparticle distance at $r^\ast=r_c^\ast \approx 2.03$, resulting in interdistance locking. Meanwhile, the angle $\alpha$ approaches a constant value, corresponding to phase locking ($\dot{\alpha}=0$). Combining these two locking conditions with Eqs.~\ref{eq:dry_r} and~\ref{eq:dry_alpha} gives
\begin{align}
C_r \leq C_m {r_c^\ast}^{p-4}\left(1+3\cos2\alpha\right),
\label{eq:locking1}
\\
C_m\sin2\alpha
= \pi {r_c^\ast}^5-\frac{C_t}{2}{r_c^\ast}^{5-q}.
\label{eq:locking2}
\end{align}
Defining
$C_{m0}=\pi {r_c^\ast}^5-\frac{C_t}{2}{r_c^\ast}^{5-q}$,
Eq.~\ref{eq:locking2} gives $\sin2\alpha=C_{m0}/C_m$, so phase locking requires $C_m\geq C_{m0}$. Linear stability analysis of Eq.~\ref{eq:dry_alpha} around the locked angle gives a relaxation rate proportional to $-\cos2\alpha$, showing that stable phase locking requires $\cos2\alpha>0$. The stable phase-locked angle therefore satisfies
$\cos2\alpha=\sqrt{1-\left(C_{m0}/{C_m}\right)^2}$.
Substituting this relation into Eq.~\ref{eq:locking1} gives
$C_r \leq {r_c^\ast}^{p-4}
\left[
C_m+3\sqrt{C_m^2-C_{m0}^2}
\right]$.
These conditions define the domain of \emph{Mode~I} [Fig.~\ref{fig1}(c)]. The condition $C_m\geq C_{m0}$ requires the transverse magnetic interaction to be strong enough to lock the orientation of the pair relative to the rotating field. At the phase-locking threshold, $C_m=C_{m0}$, the critical locked angle is $\alpha_c=45^\circ$, where the transverse magnetic interaction is maximal. The radial condition further requires the magnetic attraction, evaluated at the locked angle, to hold the particles in contact against the additional radial repulsion.

\begin{figure}
\includegraphics[width=1\linewidth]{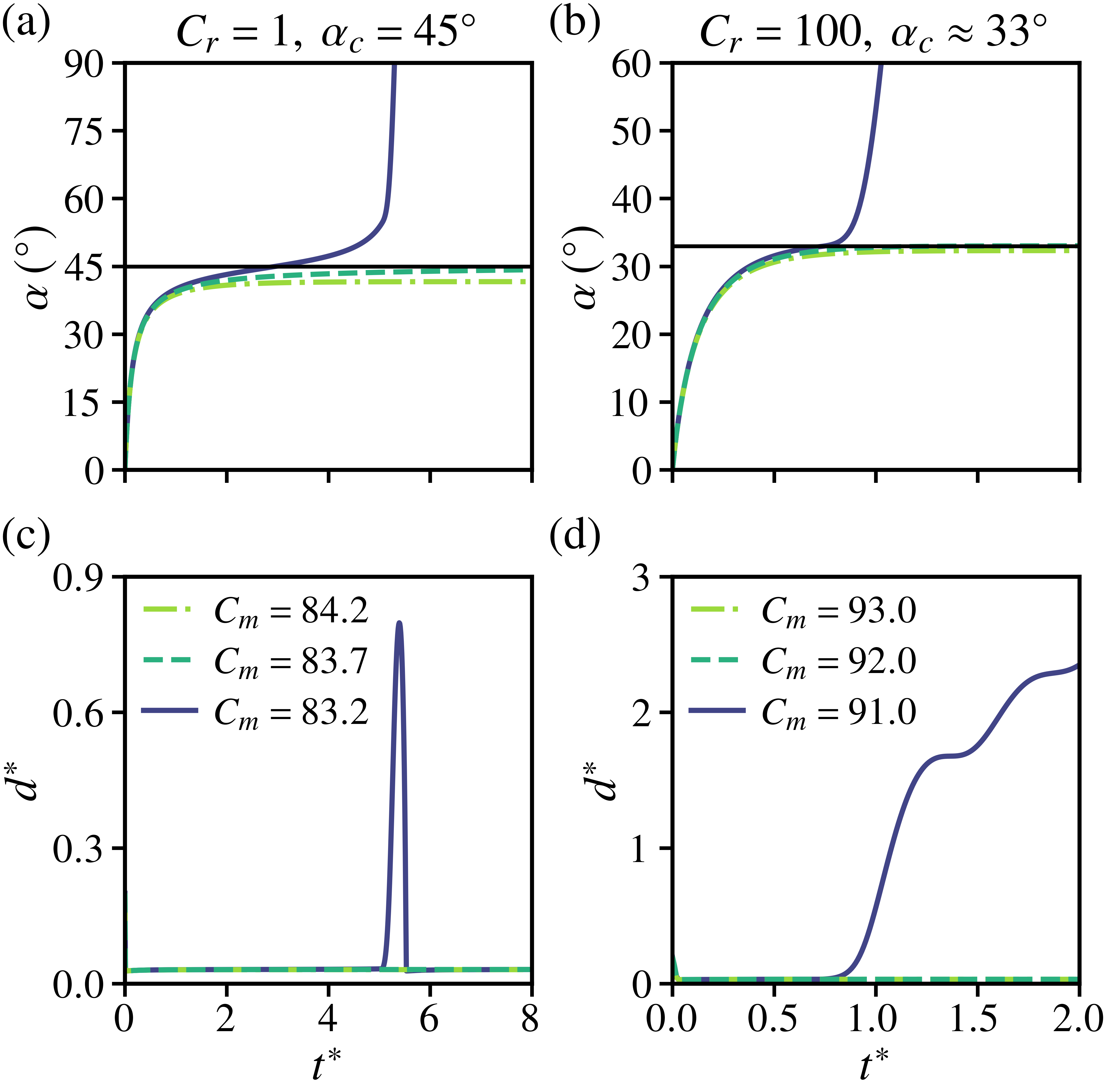}
\caption{\label{fig2}
Pair dynamics near the boundary of \emph{Mode~I} for $p=3$ and $q=5$.
(a),(b) Time evolution of the orientation angle $\alpha$ for different $C_m$ at $C_r=1$ and $C_r=100$, respectively.
(c),(d) Time evolution of the dimensionless surface-to-surface separation $d^\ast=r^\ast-2$ at $C_r=1$ and $C_r=100$, respectively.
}
\end{figure}

For weak additional radial repulsion, $C_r \leq C_{m0}{r_c^\ast}^{p-4}$, the onset of rigid-body rotation is controlled by the phase-locking threshold. For example, for p=3 and q=5, we obtain $C_{m0}\approx83.3$, so that case $C_r=1$ falls within this weak-repulsion regime. Above $C_{m0}$, $\alpha$ relaxes toward a stable locked angle, which approaches $45^\circ$ at the onset of locking, while the pair remains in contact [Figs.~\ref{fig2}(a) and~\ref{fig2}(c)]. Below $C_{m0}$, phase locking fails; $\alpha$ continues to advance, the radial magnetic interaction alternates between attraction and repulsion, and the pair enters the contact-separation rotation state.

For strong additional radial repulsion, $C_r>C_{m0}{r_c^\ast}^{p-4}$, the onset of rigid-body rotation is controlled by interdistance locking. The critical locked angle is selected by the radial balance at contact rather than by the maximum transverse magnetic interaction:
$C_r = {r_c^\ast}^{p-4} C_{m0}
\left(
1+3\cos 2\alpha_c \right)/{\sin 2\alpha_c}$.
The critical magnetic strength is then
$C_m^c=C_{m0}/{\sin 2\alpha_c}$.
For $p=3$, $q=5$, and $C_r=100$, we obtain $\alpha_c\approx33^\circ$ and $C_m^c\approx91.2$. Above this value, the pair remains locked and in contact [Figs.~\ref{fig2}(b) and~\ref{fig2}(d)]. Below it, radial repulsion overcomes the net magnetic attraction after detachment, leading to \emph{Mode~III}.

\begin{figure}
\includegraphics[width=1\linewidth]{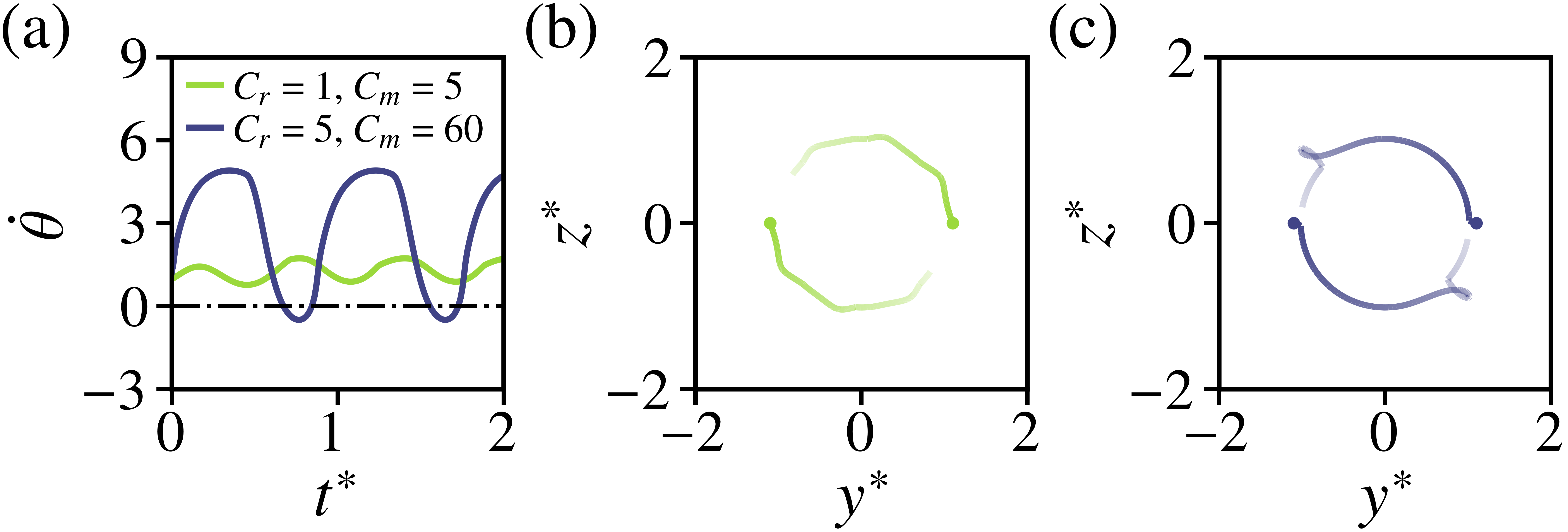}
\caption{\label{fig3}
Reversed and non-reversed orbital motion in the contact-separation rotation state.
(a) Time evolution of the rotation rate $\dot{\theta}$ of the line of centers. Negative $\dot{\theta}$ indicates local reversal of the orbital motion.
(b) Particle trajectories without reversed orbital motion for $C_r=1$ and $C_m=5$.
(c) Particle trajectories with reversed orbital motion for $C_r=5$ and $C_m=60$, showing local backtracking segments.
}
\end{figure}

In the contact-separation rotation state, the angle $\alpha$ is not locked but varies periodically. Starting from $\alpha=0$, the magnetic transverse interaction drives an increase of $\alpha$. The radial magnetic interaction changes sign when
$\alpha=0.5\arccos\left(-1/3\right)\approx54.74^\circ$.
Thus, during one rotation cycle, the radial magnetic interaction switches between attraction and repulsion, producing repeated contact and separation.

The transverse dynamics further determines whether the orbital motion undergoes local reversal. We define $\theta=2\pi t^\ast-\alpha$ as the orientation angle of the line of centers. Reversed orbital motion occurs when $\dot{\theta}<0$. Using Eq.~\ref{eq:dry_alpha}, this gives
$C_t/r^{\ast q}+
2C_m\sin2\alpha/r^{\ast5}<0$ . 
For $p=3$ and $q=5$, the reversal condition becomes independent of $r^\ast$, reducing to $C_t+2C_m\sin2\alpha<0$. The strongest retardation occurs at $\sin2\alpha=-1$, yielding the reversal threshold $C_m^{\rm rev}=C_t/2$. For $C_t=50$, this gives $C_m^{\rm rev}=25$. When $C_m>C_m^{\rm rev}$, $\dot{\theta}$ becomes negative during part of the cycle [Fig.~\ref{fig3}(a)], and the particle trajectories develop local backtracking segments [Fig.~\ref{fig3}(c)]. When $C_m<C_m^{\rm rev}$, $\dot{\theta}$ remains positive [Fig.~\ref{fig3}(a)] and the trajectory rotates monotonically [Fig.~\ref{fig3}(b)].

The boundary between \emph{Mode~II} and \emph{Mode~III} can be estimated by averaging the radial dynamics over one cycle of the angle $\alpha$, assuming that $r^\ast$ changes weakly during this cycle. The magnetic radial interaction then gives a net attractive contribution,
$\left\langle -C_m(1+3\cos2\alpha)/r^{\ast4} \right\rangle
\simeq -C_m/r^{\ast4}$.
Balancing this averaged magnetic attraction with the additional radial repulsion gives a fixed point
$\bar r_0^\ast=\left(C_m/C_r\right)^{1/(4-p)}$.
Because we focus on long-range radial repulsion that decays more slowly than the magnetic dipolar interaction, we mainly consider $p<4$. In this case, the fixed point is unstable. If the initial separation satisfies $r^\ast(0)<\bar r_0^\ast$, the averaged radial drift is inward and the particles return to contact, leading to \emph{Mode~II}. If $r^\ast(0)>\bar r_0^\ast$, the averaged drift is outward and the particles separate, leading to \emph{Mode~III}. The transition is therefore estimated by $r^\ast(0)=\bar r_0^\ast$, which gives
$C_m=C_r r^\ast(0)^{4-p}$.
The resulting boundary agrees well with the numerically observed boundary between \emph{Mode~II} and \emph{Mode~III} [Fig.~\ref{fig1}(c) and Fig.~S2 of the Supplemental Material].

\begin{figure}
\includegraphics[width=0.9\linewidth]{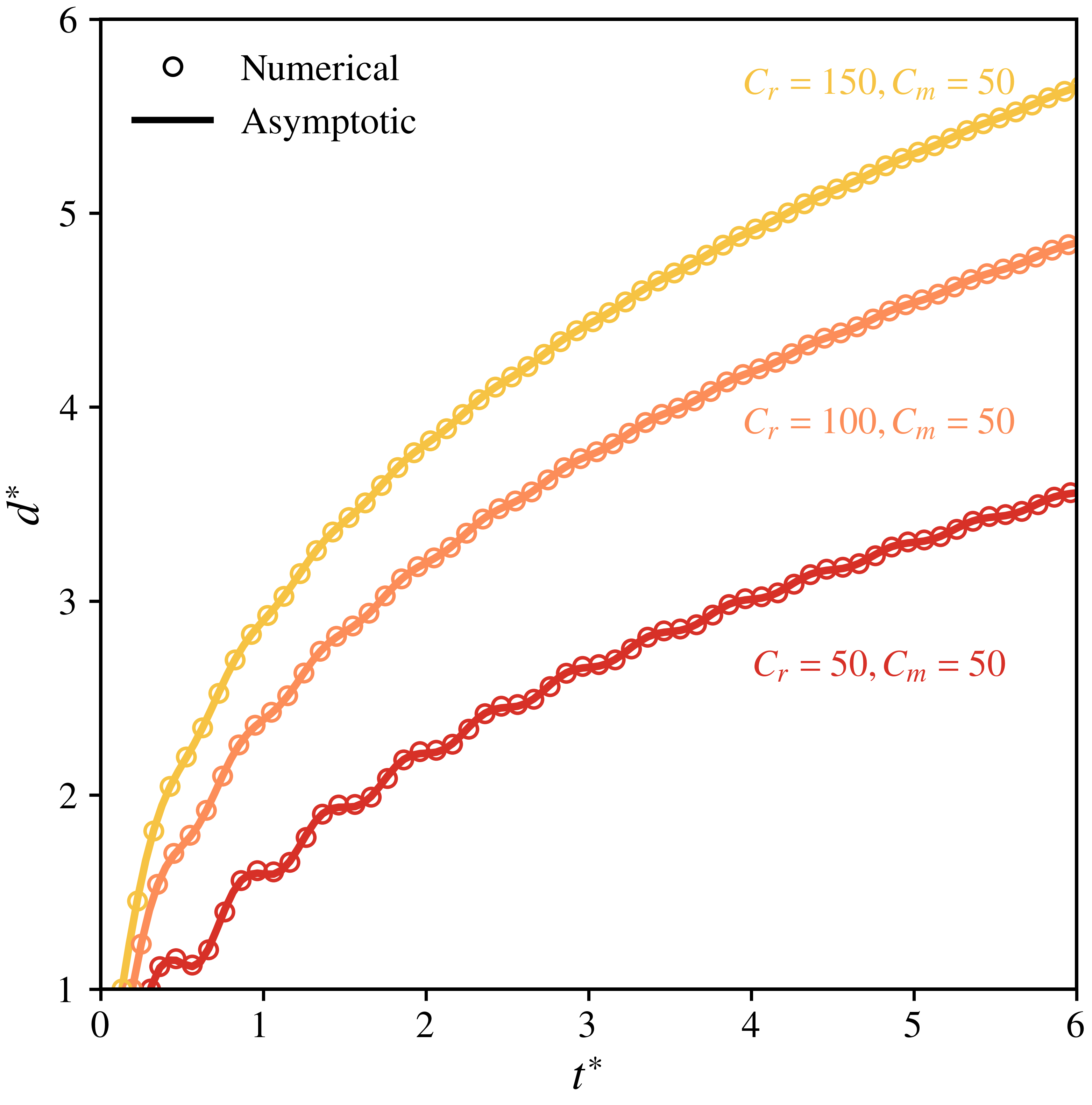}
\caption{\label{fig4}
Asymptotic solution for particle separation in the separation rotation state.
The surface-to-surface gap $d^\ast$ is plotted as a function of $t^\ast$ for different strengths of the additional radial repulsion. Symbols show numerical integration of the full reduced system, and solid lines show the corresponding asymptotic solution.
}
\end{figure}

When the particles separate, the interparticle distance becomes large, $r^\ast\gg 1$. Because the transverse coupling in Eq.~\ref{eq:dry_alpha} decays rapidly with distance, the angle can be approximated as $\alpha(t^\ast)\simeq 2\pi(t^\ast-t_1^\ast)+\alpha_1$, where $r^\ast(t_1^\ast)=3$ and $\alpha_1=\alpha(t_1^\ast)$. Equation~\ref{eq:dry_r} then reduces to a slow radial evolution driven by the cycle-averaged repulsion, with a small oscillatory correction from the magnetic interaction. This separation of time scales allows a multiscale expansion, detailed in the Supplemental Material, which yields the leading-order approximation
\begin{equation}
\begin{aligned}
r^\ast(t^\ast)&\approx \Bigl[
f_0(t^\ast-t_1^\ast)\\
&-15\varepsilon C_m
\sin\!\bigl(4\pi(t^\ast-t_1^\ast)+2\alpha_1\bigr)+\varepsilon f_1(t^\ast-t_1^\ast)
\Bigr]^{1/5},
\end{aligned}
\label{eq:modeIII_asymp}
\end{equation}
where $\varepsilon=1/4\pi$. This expression captures the separation dynamics in \emph{Mode~III}: a slow increase of the mean interparticle distance modulated by small magnetic oscillations. The asymptotic solution agrees well with the numerical integration of the full reduced model [Fig.~\ref{fig4}].

To examine a specific hydrodynamic realization of the additional radial repulsion and transverse coupling in the reduced model, we perform lattice Boltzmann simulations using Ludwig, an open-source package for complex-fluid simulations~\cite{Ludwig2026}. The numerical system consists of two spherical particles of radius $R$ placed on a flat wall, with the wall-normal direction along $x$. The particles are immersed in a Newtonian fluid of density $\rho$ and dynamic viscosity $\eta$. No-slip boundary conditions are imposed on both the particle surfaces and the solid wall, while periodic boundary conditions are applied in the $y$ and $z$ directions. The magnetic interaction between the particles is applied according to Eq.~\ref{eq:mag}. Each particle is driven to rotate synchronously by an external magnetic field at angular frequency $\omega$, with the rotation axis perpendicular to the wall. The Reynolds number is defined as $\mathrm{Re}= \rho \omega R^2 / \eta$.

This setup provides one physical realization of the reduced model. At $\mathrm{Re} = O(1)$, particle rotation generates an azimuthal flow that contributes to the transverse coupling, as well as an inertial secondary flow that is directed inward near the polar regions and outward near the equatorial region, giving rise to an effective radial repulsion between the two rotating particles~\cite{Bickley1938,ShenLintuvuori2020}. Thus, in this example, the additional radial interaction and transverse coupling both arise from the flow generated by particle rotation.

A quantitative mapping from the hydrodynamic simulations to the reduced coefficients is not straightforward, because the flow contains finite-inertia effects, wall-mediated hydrodynamic interactions, and near-field corrections. Nevertheless, the far-field decay and the dependence on control parameters provide useful estimates. Owing to the presence of the wall, the azimuthal flow generated by a rotating particle decays approximately as $u_\phi^\ast\sim r^{\ast-4}$ in the far field~\cite{BlakeChwang1974,DeanONeill1963,LiuProsperetti2010}, as confirmed by fitting the simulation data in Fig.~S1 of the Supplementary Material. Since the angular drift of the line of centers scales as $u_\phi^\ast/r^\ast$, this gives $q=5$ in the reduced model.

The inertial secondary flow generates an outward radial velocity near the equatorial region. Its far-field magnitude scales as $u_r^\ast\sim \mathrm{Re}\,r^{\ast-3}$, consistent with the fitting results in Fig.~S1 of the Supplemental Material. This scaling gives the radial exponent $p=3$ and indicates that the strength of the additional radial repulsion scales as $C_r\sim \mathrm{Re}$. The magnetic coefficient is estimated by comparing the magnetic dipolar force with the effective viscous resistance. To account phenomenologically for finite-inertia, wall-mediated, and near-field corrections~\cite{ONeill1964,Goldman1967}, we write the resistance correction factor as $\chi(1+\zeta\mathrm{Re})$, where $\chi$ accounts for the wall-induced modification of the viscous resistance and $\zeta$ is a fitting parameter accounting for the finite-inertia correction. Estimates of $\chi$ and $\zeta$ are provided in the Supplemental Material. The dimensionless magnetic forcing parameter is therefore defined as $C_m=\mu_0m^2/[\chi(1+\zeta\mathrm{Re})\eta\omega_0R^6]$.

\begin{figure}
\includegraphics[width=1\linewidth]{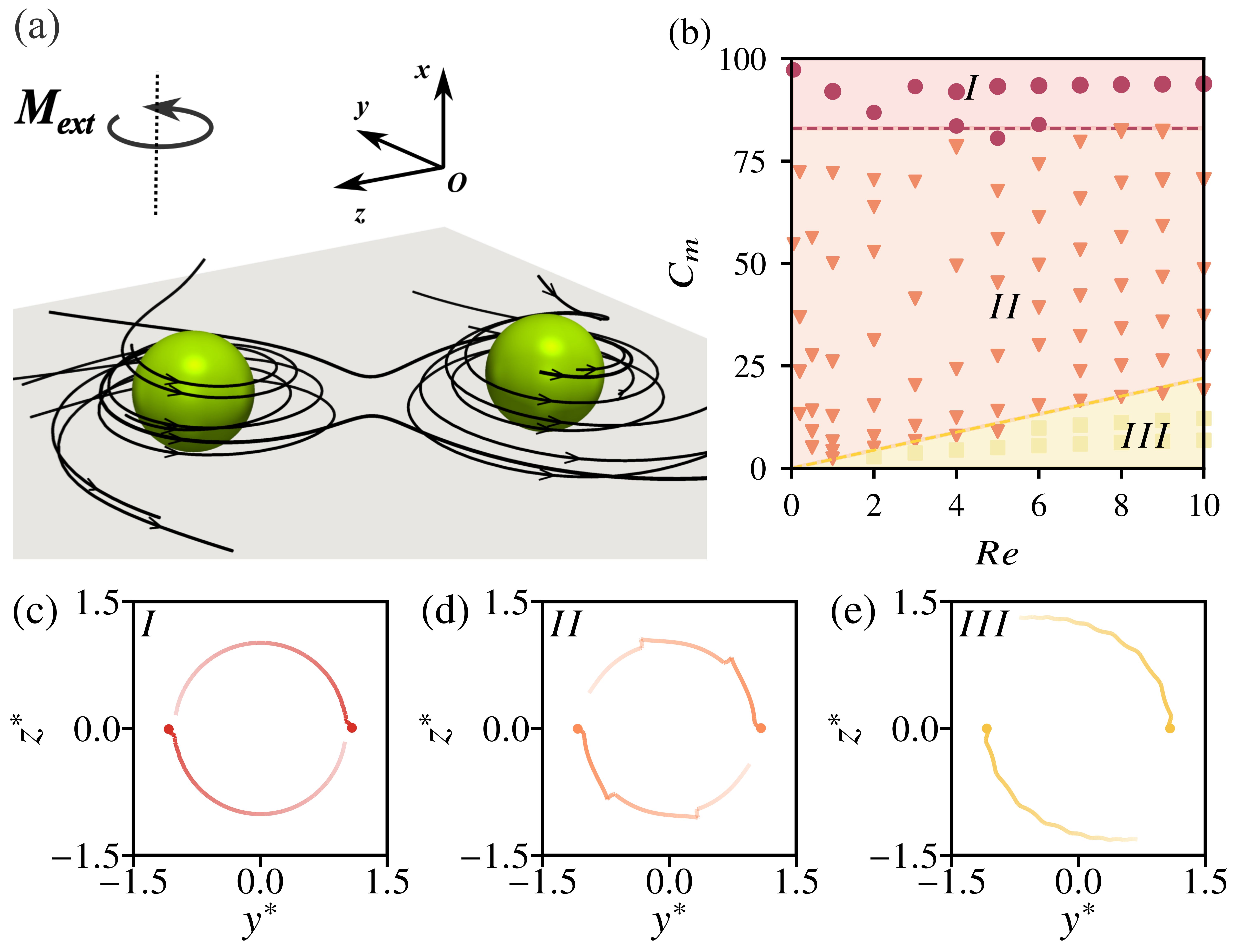}
\caption{\label{fig5}
Hydrodynamic simulations of two synchronized magnetic particles rotating near a wall.
(a) Schematic of the simulation setup, showing the near-wall flow generated by two rotating particles driven by an external magnetic field $\boldsymbol{M}_{\rm ext}$ at $\mathrm{Re}\sim 10$.
(b) Phase diagram of pair dynamics in the $(\mathrm{Re},C_m)$ parameter space. \emph{Mode~I}: rigid-body rotation; \emph{Mode~II}: contact-separation rotation; \emph{Mode~III}: irreversible separation.
(c)–(e) Representative particle trajectories for \emph{Mode~I}–\emph{Mode~III}, respectively.
}
\end{figure}

These estimates provide a straightforward mapping from the reduced $(C_r,C_m)$ parameter space to the physical $(\mathrm{Re},C_m)$ plane, with $C_r\sim \mathrm{Re}$. The lattice Boltzmann phase diagram [Fig.~\ref{fig5}(b)] reproduces the three-mode structure of the reduced model [Fig.~\ref{fig1}(c)]: rigid-body rotation, contact-separation rotation, and irreversible separation [Figs.~\ref{fig5}(c)–~\ref{fig5}(e)]. Because $\mathrm{Re}<10$ in the present simulations, only relatively small $C_r$ values over a limited range is explored.

The transition trends also follow the reduced-model predictions. The \emph{Mode~I}-\emph{Mode~II} boundary is expected to occur at approximately constant $C_m$ when $C_r$ is small, because it is controlled mainly by the phase-locking threshold. The \emph{Mode~II}-\emph{Mode~III} boundary is expected to scale as $C_m\sim \mathrm{Re}$, reflecting the balance between magnetic attraction and the inertia-induced radial repulsion. Both trends are consistent with the lattice Boltzmann results [Fig.~\ref{fig5}(b)]. This agreement shows that the reduced model captures the leading pair dynamics of rotating magnetic dipoles with an additional radial repulsion and rotation-induced transverse coupling.

In summary, we have shown that the pair motion of synchronized rotating magnetic microparticles is selected by the competition between radial and transverse interactions. Radial interactions regulate contact, separation, and recontact, whereas transverse coupling controls the orientation of the pair relative to the rotating field. These balances organize the observed trajectories into predictable motion modes and yield explicit transition criteria for switching between them.

These criteria provide design rules for programming elementary pair dynamics. By tuning magnetic attraction against radial repulsion and transverse coupling, a rotating pair can be selected to remain bound, undergo contact-separation motion, or separate irreversibly. The hydrodynamic simulations further show that such effective interactions can arise naturally from rotational flows and finite-inertia secondary flows. The present framework therefore links pair-level interaction design to controllable motion modes, offering a basis for organizing larger assemblies of rotating magnetic particles in field-driven colloidal and microrobotic systems.

We thank Jinhan Xie for useful discussions.
Z.S. and D.F. acknowledge the Natural Science Foundation of Beijing, China (Grant No. 1252020) for funding. X.Z. and L.W. acknowledge the National Key R\&D Program of China (2022YFF0503504).

\clearpage
\onecolumngrid
\setcounter{section}{0}
\setcounter{subsection}{0}
\setcounter{figure}{0}
\setcounter{table}{0}
\setcounter{equation}{0}
\renewcommand{\thefigure}{S\arabic{figure}}
\renewcommand{\theequation}{S\arabic{equation}}
\begin{center}
{\large\bfseries Supplementary Material for\\[0.4em]
``Controlling Pair Dynamics of Rotating Magnetic Microparticles through Radial and Transverse Interactions''}
\end{center}
\vspace{1em}

\section{Hydrodynamic coupling and force estimate}

A steadily rotating sphere generates an azimuthal flow and, at finite $\mathrm{Re}$, an inertia-induced secondary flow. The latter produces an outward radial flow in the equatorial plane and provides the physical origin of the flow-induced radial repulsion introduced in the reduced model. For $\mathrm{Re}=\rho \omega R^2/\eta<O(1)$, Bickley’s asymptotic solution ~\cite{Bickley1938} for an isolated sphere rotating with angular speed $\omega$ is given in spherical coordinates $(r,\phi,\psi)$ by
\begin{align}
\boldsymbol{u}_{r}(r) &=
\left[
-\frac{\omega R^{3}}{8r^{2}}
(3\cos^{2}\psi - 1)
\left(1 - \frac{R}{r}\right)^{2}
\mathrm{Re}
\right]\boldsymbol{\hat{e}}_{r},
\label{eq:bickley_vr}
\\[4pt]
\boldsymbol{u}_{\phi}(r) &=
\left[
\frac{\omega R^{3}}{r^{2}}\sin\psi + O(\mathrm{Re}^{2})
\right]\boldsymbol{\hat{e}}_{\phi},
\label{eq:bickley_vtheta}
\\[4pt]
\boldsymbol{u}_{\psi}(r) &=
\left[
\frac{\omega R^{4}}{4r^{3}}
\left(1 - \frac{R}{r}\right)
\sin\psi\cos\psi\,\mathrm{Re}
\right]\boldsymbol{\hat{e}}_{\psi}.
\label{eq:bickley_vpsi}
\end{align}
In the equatorial plane, Eq.~\eqref{eq:bickley_vr} predicts an outward radial flow, whereas Eq.~\eqref{eq:bickley_vtheta} describes the azimuthal flow responsible for the transverse advection between rotating particles. These expressions are derived for an isolated sphere in an unbounded fluid. A nearby no-slip wall modifies both the flow amplitudes and their far-field decay, while preserving the same physical roles of the azimuthal and secondary flows.

The modification induced by the wall can be understood from the image system of the low-Reynolds-number singularities~\cite{BlakeChwang1974}. In an unbounded fluid, the leading azimuthal flow generated by a rotating sphere is a rotlet with $u_\phi \sim r^{-2}$. For a rotlet oriented normal to a no-slip wall, the image rotlet has the opposite sign. If both the source and the observation point are located at the same height $h$ above the wall and separated laterally by $r\gg h$, the combined contribution from the real and image rotlets becomes
\begin{align}
u_\phi(r)
&\sim
\frac{1}{r^2}
-\frac{r}{(r^2+4h^2)^{3/2}}
\nonumber\\
&=\frac{6h^2}{r^4}+O(r^{-6}).
\label{eq:wall_rotlet_decay}
\end{align}
Therefore, the $r^{-2}$ contribution cancels in the equal-height plane, and the leading azimuthal flow decays as $r^{-4}$. This far-field scaling agrees with the exact creeping-flow solution for a sphere rotating near a plane wall derived by Dean and O’Neill~\cite{DeanONeill1963}.

\begin{figure*}
\includegraphics[width=0.99\linewidth]{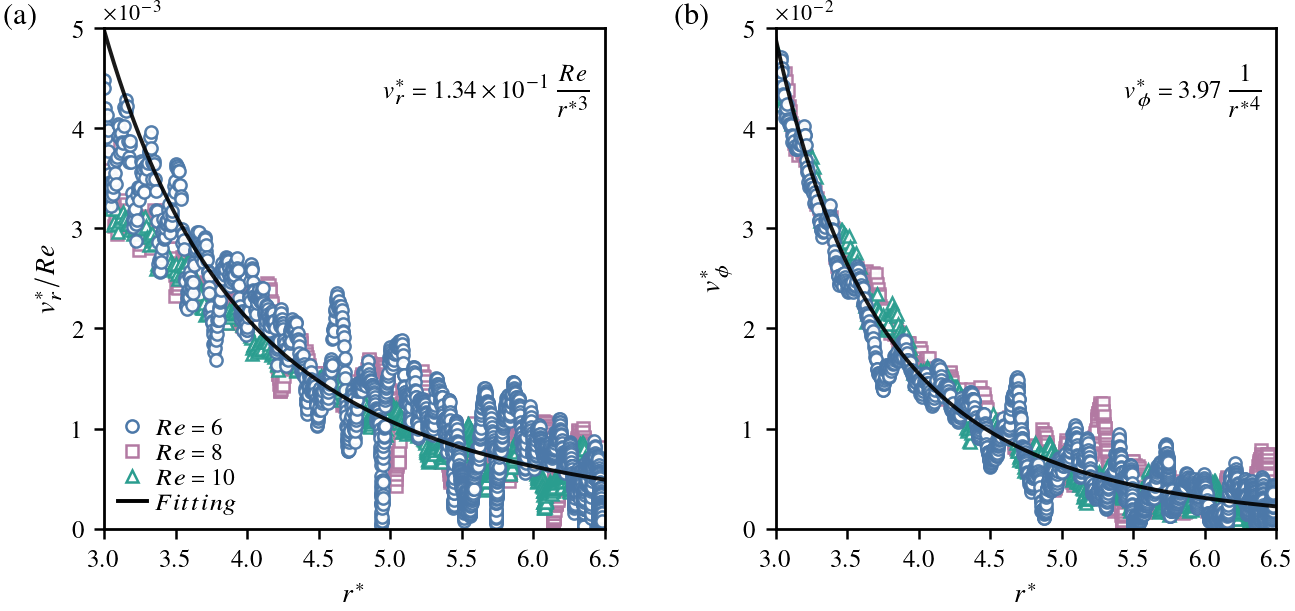}
\caption{\label{fig:hydro_scaling}
Dimensionless particle velocities obtained from lattice Boltzmann simulations for two rotating particles interacting hydrodynamically near a no-slip wall. The velocities are normalized by $\omega R$. (a) Radial velocity $v_r^*$. (b) Azimuthal velocity $v_\phi^*$.
}
\end{figure*}

The inertia-induced radial flow in Eq.~\eqref{eq:bickley_vr} has a stresslet-like far-field structure in an unbounded fluid. For singularities with stresslet-type far fields, the image system associated with a no-slip wall suppresses the leading normal contribution, so that the first non-vanishing far-field term decays one power faster with distance. Consequently, the leading radial flow is expected to decay as $r^{-3}$ instead of the free-space $r^{-2}$ behavior. Since the secondary flow itself is already $O(\mathrm{Re})$ relative to the characteristic azimuthal velocity $\omega R$, this gives the far-field scaling $u_r\sim \mathrm{Re}\,r^{-3}$.
This scaling is a far-field estimate based on the wall-modified multipole structure rather than a rigorous finite-$\mathrm{Re}$ asymptotic solution.

We measured the translational velocities of two rotating particles interacting through the flows they generate near a no-slip wall in lattice Boltzmann simulations. The dimensionless radial and transverse velocities, $v_r^*=v_r/\omega R$ and $v_\phi^*=v_\phi/\omega R$, shown in Fig.~\ref{fig:hydro_scaling}, are well fitted by
\begin{align}
v_r^* &= 0.134\frac{\mathrm{Re}}{r^{*3}},
\label{eq:fit_vr}
\\
v_\phi^* &= 3.97\frac{1}{r^{*4}}.
\label{eq:fit_vphi}
\end{align}

Here, the hydrodynamic interaction between the particles is dominated by advection in the flow generated by the other rotating particle. The measured particle velocities therefore reflect the radial and azimuthal components of the particle-induced flow. The fitted decay exponents agree with the far-field scalings predicted from the wall-modified multipole arguments above. Moreover, the collapse of $v_r^*/\mathrm{Re}$ for $Re =$ 6, 8, and 10 confirms the leading linear dependence of the secondary radial flow on $\mathrm{Re}$. These results indicate that the hydrodynamic interaction between two rotating
particles near a no-slip wall at finite Reynolds number is well approximated by $p= 3$ in the radial direction and $q= 5 $ in the transverse direction. Here, $q= 5 $ arises because the angular velocity of the line joining the particle centers scales as $\dot{\alpha} \sim v_\phi ^*/r^*$. These values are therefore used in the reduced model to compare its predictions with the lattice Boltzmann simulations.

To estimate the transverse coupling strength $C_t$ in the reduced model, we
measure the transverse particle velocity in two-particle lattice Boltzmann
simulations. Since both particles rotate and experience equal transverse
advection, their relative transverse velocity is $2v_\phi$. For small angular
displacements, the angular velocity of the line joining the particle centers
is given by the relative transverse velocity divided by the particle
separation, $d\theta/dt=2v_\phi/r$. The hydrodynamic contribution to
the angular velocity therefore becomes
\begin{align}
\left.\frac{d\theta}{dt^*}\right|_{\rm h}
&=
\frac{2\pi}{\omega}\frac{2v_\phi}{r}
=\frac{4\pi v_\phi^*}{r^*}
=\frac{4\pi(3.97)}{r^{*5}}.
\label{eq:relative_transverse_mapping}
\end{align}
Since $\alpha=2\pi t^*-\theta$, the hydrodynamic contribution enters the
equation for $\alpha$ with a negative sign. Comparing the above expression
with the reduced model yields
\begin{align}
q&=5,
&
C_t&=4\pi(3.97)=49.89\simeq50.
\label{eq:Ct_mapping}
\end{align}

To estimate the effective hydrodynamic resistance in the reduced model, we
account for finite-inertia, wall-mediated, and near-field effects. The
effective resistance coefficient is written as
$6\pi\eta R\,\chi(1+\zeta\mathrm{Re})$, where the Stokes-limit factor $\chi$
is taken from O'Neill's solution for a sphere translating parallel to a
plane wall~\cite{ONeill1964}. For the simulated center-to-wall distance,
$d/R=1.03333$, this gives $\chi\simeq2.8$.

The remaining factor $(1+\zeta\mathrm{Re})$ provides a phenomenological
correction for finite-inertia effects. To determine $\zeta$, we use the
Mode-I--Mode-II transition, for which the reduced model predicts an
approximately constant critical magnetic coefficient in the small-$C_r$
regime. Fitting the transition data with this constant critical value yields
$\zeta=0.18$. The reduced magnetic coefficient used in the main text is
therefore $C_m=\mu_0m^2/[2.8(1+0.18\mathrm{Re})\eta\omega_0R^6]$.

\clearpage

\section{Robustness to the interaction decay exponents}

The reduced pair dynamics are generalized by allowing the additional radial
repulsion and the transverse coupling to have independent algebraic decay
exponents, $p$ and $q$, respectively:

\begin{align}
\frac{dr^\ast}{dt^\ast}
&= \frac{C_r}{r^{\ast p}}
-\frac{C_m\bigl(1+3\cos 2\alpha\bigr)}{r^{\ast 4}},
\label{eq:supp_dry_r}
\\
\frac{d\alpha}{dt^\ast}
&= 2\pi
-\frac{C_t}{r^{\ast q}}
-\frac{2C_m\sin 2\alpha}{r^{\ast 5}}.
\label{eq:supp_dry_alpha}
\end{align}

Here, $C_r$, $C_m$, and $C_t$ denote the strengths of the additional radial
repulsion, magnetic dipolar interaction, and transverse coupling,
respectively. The exponents $p$ and $q$ determine the spatial decay of the
additional radial interaction and the transverse coupling, whereas the
magnetic interactions retain the point-dipole scaling.

\begin{figure*}[b]
\includegraphics[width=0.99\linewidth]{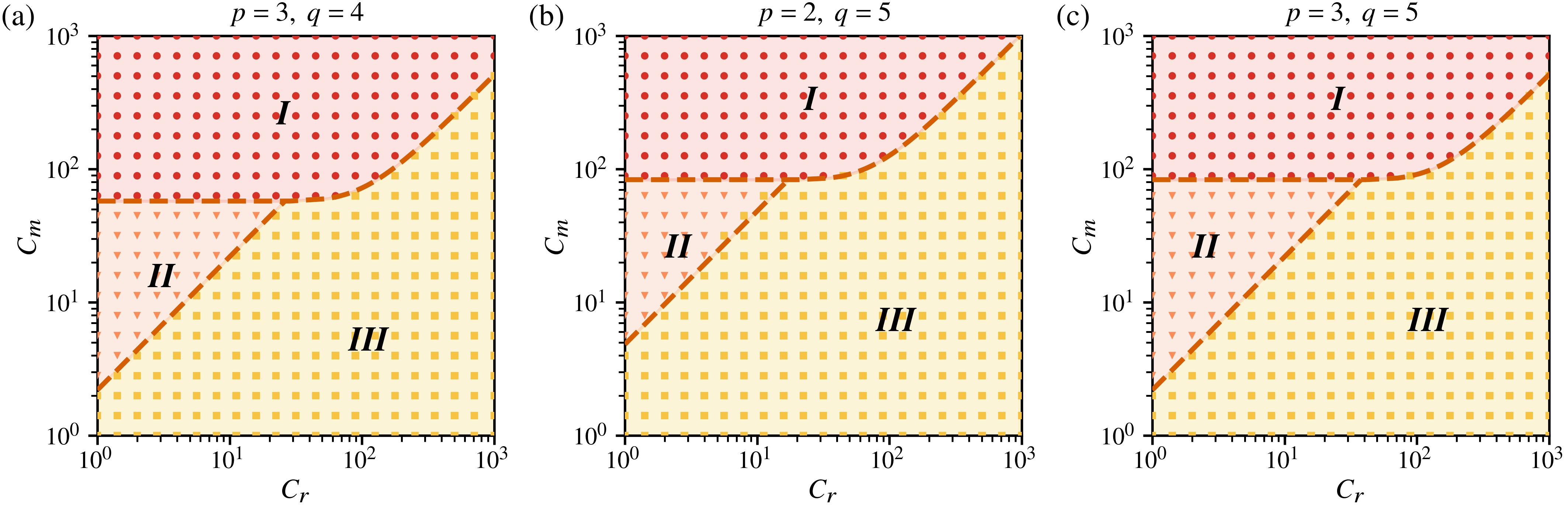}
\caption{\label{fig:phase_exponents}
Phase diagrams of the generalized reduced model in the $(C_r,C_m)$ parameter plane for different interaction decay exponents:
(a) $p=3$, $q=4$;
(b) $p=2$, $q=5$; and
(c) $p=3$, $q=5$.
Circles, downward triangles, and squares denote the numerical results for Modes I, II, and III, respectively. Dashed curves show the theoretical phase boundaries.
}
\end{figure*}

Figure~\ref{fig:phase_exponents} compares three representative combinations
of the interaction decay exponents. In all cases, numerical integration of
Eqs.~\eqref{eq:supp_dry_r} and \eqref{eq:supp_dry_alpha} produces the same three motion
modes identified in the main text: rigid-body rotation (Mode~I),
contact-separation rotation (Mode~II), and irreversible separation
(Mode~III). Changing $p$ or $q$ shifts the phase boundaries quantitatively
because it changes the relative ranges of the radial and transverse
interactions. However, the three dynamical modes and their ordering in the
phase diagram remain unchanged over the parameter combinations considered
here. The numerical phase boundaries are also in good agreement with the
corresponding theoretical predictions.

These results indicate that the three dynamical modes arise from the
competition between radial and transverse interactions and persist over a
broad range of interaction scalings. This robustness suggests that the
mode-selection mechanism is generic and does not rely on a particular choice
of interaction exponents.

\clearpage

\section{Asymptotic analysis for the separation dynamics}

This section presents the far-field asymptotic analysis used to derive the
second-order approximation for the separation dynamics in Mode~III shown in
Fig.~4 of the main text. For the interaction exponents $p=3$ and $q=5$, the
reduced model reads

\begin{align}
\frac{dr^*}{dt^*} &= \frac{ C_r}{r^{*3}} - \frac{C_m(1+3\cos 2\alpha)}{ r^{*4}}, \\
\frac{d\alpha}{dt^*} &= 2\pi - \frac{C_t }{r^{*5}}-\frac{2C_m \sin 2 \alpha}{ r^{*5}}.
\label{eq:dr_dalpha}
\end{align}

In the far-field limit, the two particles are well separated,
$r^*\gg r_c^*$,
so that

\begin{equation}
\frac{ C_t }{2\pi r^{*5}} = o(1), \quad \frac{C_m}{\pi r^{*5}} = o(1).
\label{eq:far_field}
\end{equation}

These conditions correspond to the irreversible separation dynamics in
Mode~III.

To exclude the near-field part of the trajectory from the asymptotic analysis,
we define $t_1^*$ as the time at which the separating pair first reaches
$r^*=3$, and denote the phase at that instant by $\alpha_1$:
\begin{equation}
  r^*(t_1^*)=3,
  \qquad
  \alpha_1\equiv\alpha(t_1^*).
  \label{eq:far_initial}
\end{equation}
We then introduce the elapsed time measured from this far-field entry point,
\begin{equation}
  \tau=t^*-t_1^*,
\end{equation}
so that the initial condition for the asymptotic problem is imposed at
$\tau=0$. At leading order, Eq.~\eqref{eq:supp_dry_alpha} gives
\begin{equation}
  \alpha(t^*)\simeq 2\pi(t^*-t_1^*)+\alpha_1
  =2\pi\tau+\alpha_1.
  \label{eq:alpha_far}
\end{equation}
Because $d/dt^*=d/d\tau$, Eq.~\eqref{eq:supp_dry_r} therefore reduces to
\begin{align}
\frac{dr^*}{d\tau}
&= \frac{C_r}{r^{*3}}
- \frac{C_m\bigl[1+3\cos(4\pi\tau+2\alpha_1)\bigr]}{r^{*4}}.
\label{eq:dr_dalpha_far}
\end{align}

Since this equation involves distinct fast and slow timescales, we introduce
$\varepsilon=1/(4\pi)$, define the fast phase
\begin{equation}
  T=\frac{\tau}{\varepsilon}+2\alpha_1
  =4\pi\tau+2\alpha_1,
\end{equation}
and set $s=r^{*5}$. Regarding $s=s(\tau,T)$, the chain rule gives
\begin{equation}
  \frac{ds}{d\tau}
  = \frac{1}{\varepsilon}\,\frac{\partial s}{\partial T}
  + \frac{\partial s}{\partial \tau}
  = 5C_r s^{1/5} - 5C_m\bigl(1+3\cos T\bigr),
  \qquad \varepsilon = \frac{1}{4\pi}.
  \label{eq:s-eq-1}
\end{equation}

The corresponding asymptotic expansion is
\begin{equation}
  s(\tau,T)
  = s_0(\tau,T) + \varepsilon s_1(\tau,T)
    + \varepsilon^2 s_2(\tau,T) + \cdots,
  \label{eq:s-expansion}
\end{equation}
with the initial condition
\begin{equation}
  s(0,2\alpha_1)=3^5.
  \label{eq:s-initial}
\end{equation}

Substituting the above expressions into
\eqref{eq:s-eq-1} and collecting like powers of $\varepsilon$, we find
\begin{align}
  &O(\varepsilon^{-1}):\qquad
    \partial_T s_0 = 0,
    \label{eq:order-minus1}
  \\
  &O(\varepsilon^{0}):\qquad
    \partial_T s_1 + \partial_{\tau} s_0
    = 5C_r s_0^{1/5} - 5C_m\bigl(1+3\cos T\bigr),
    \label{eq:order-0}
  \\
  &O(\varepsilon^{1}):\qquad
    \partial_T s_2 + \partial_{\tau} s_1
    = C_r s_0^{-4/5} s_1.
    \label{eq:order-1}
\end{align}

From \eqref{eq:order-minus1} we conclude that $s_0$ is independent of
$T$ and thus depends only on $\tau$:
\begin{equation}
  s_0 = f_0(\tau).
\end{equation}

Eliminating the secular term in \eqref{eq:order-0} yields the solvability condition
\begin{equation}
  \frac{df_0}{d\tau}
  = 5C_r f_0^{1/5} - 5C_m,
  \qquad f_0(0)=3^5.
  \label{eq:s0-eq}
\end{equation}
This gives the following implicit equation for $f_0(\tau)$:
\begin{equation}
\frac{f_0^{4/5}}{4C_r} + \frac{C_m f_0^{3/5}}{3C_r^2} + \frac{C_m^2 f_0^{2/5}}{2C_r^3} + \frac{C_m^3 f_0^{1/5}}{C_r^4} 
+ \frac{C_m^4}{C_r^5} \ln \left| C_r f_0^{1/5} - C_m \right| = \tau + C,
  \label{eq:f0-implicit}
\end{equation}
where the constant $C$ is determined by $f_0(0)=3^5$:
\begin{equation}
C = \frac{3^4}{4C_r} + \frac{3^3 C_m}{3C_r^2} + \frac{3^2 C_m^2}{2C_r^3} 
+ \frac{3 C_m^3}{C_r^4} + \frac{C_m^4}{C_r^5} \ln |3C_r - C_m|.
  \label{eq:C-implicit}
\end{equation}

The oscillatory correction satisfies
\begin{equation}
  \partial_T s_1
  = -15 C_m \cos T.
\end{equation}
Integrating with respect to $T$ yields
\begin{equation}
  s_1(\tau,T)
  = -15C_m \sin T + f_1(\tau).
  \label{eq:s1-sol}
\end{equation}
The $O(\varepsilon)$ initial condition is
$s_1(0,2\alpha_1)=0$, and hence
\begin{equation}
  f_1(0)=15C_m\sin(2\alpha_1).
  \label{eq:f1-initial}
\end{equation}

At order $O(\varepsilon)$, we have
\begin{equation}
  \partial_T s_2 + \partial_{\tau} f_1
  = C_r f_0^{-4/5}\bigl(-15C_m \sin T + f_1(\tau)\bigr).
\end{equation}

The corresponding solvability condition is
\begin{equation}
  \frac{df_1}{d\tau}
  = C_r f_0^{-4/5} f_1,
  \qquad f_1(0)=15C_m\sin(2\alpha_1).
  \label{eq:f1-eq}
\end{equation}
Once $f_0(\tau)$ is known, this equation gives
\begin{equation}
  \begin{aligned}
  f_1(\tau)
  ={}& 15C_m \sin(2\alpha_1) \\
  &\times \exp\left[
    C_r \int_{0}^{\tau} f_0(\xi)^{-4/5}\,d\xi
  \right].
  \end{aligned}
\end{equation}

In summary, the multiple-scale expansion gives
\begin{equation}
  s(\tau,T)
  = f_0(\tau) - 15\varepsilon C_m \sin T + \varepsilon f_1(\tau)
    + O(\varepsilon^2),
  \label{eq:s-perturbation}
\end{equation}

Returning to the original time $t^*$, with $\tau=t^*-t_1^*$, we obtain
\begin{equation}
  r^*(t^*)
  = \left[
      f_0(\tau)
      - \frac{15 C_m}{4\pi}\sin\bigl(4\pi\tau+2\alpha_1\bigr)
      + \frac{1}{4\pi}f_1(\tau)
    \right]^{1/5},
  \qquad \tau=t^*-t_1^*.
  \label{eq:r-perturbation}
\end{equation}
which fits well with the numerical results in Mode~III, as shown in Fig.~4.

\end{document}